\documentclass[letterpaper]{article} 
\usepackage{aaai25} 
\usepackage{times} 
\usepackage{helvet} 
\usepackage{courier} 
\usepackage[hyphens]{url} 
\usepackage{graphicx} 
\usepackage{natbib} 
\usepackage{caption} 
\usepackage{newtxtext}
\usepackage{newtxmath}

\usepackage{makecell}
\usepackage{multicol}
\usepackage{multirow}

\newcommand{\eg}{\textit{e}.\textit{g}.}

\title{Learning Generalized Residual Exchange-Correlation-Uncertain Functional for Density Functional Theory}

\author{
   Sizhuo Jin\textsuperscript{\rm 1}, 
   Shuo Chen\textsuperscript{\rm 2}, 
   Jianjun Qian\textsuperscript{\rm 1}, 
   Ying Tai\textsuperscript{\rm 2}, 
   Jun Li\textsuperscript{\rm 1}\thanks{corresponding author}
}
\affiliations{
    \textsuperscript{\rm 1}School of Computer Science and Engineering, Nanjing University of Science and Technology, Nanjing, China. \\

    \textsuperscript{\rm 2}School of Intelligence Science and Technology, Nanjing University, Suzhou, China. \\
    
    \{jinsizhuo, csjqian, junli\}@njust.edu.cn, \ \ shuo.chen.ya@foxmail.com, \ \ yingtai@nju.edu.cn

}

\begin{document}
\maketitle
\begin{abstract}
    Density Functional Theory (DFT) stands as a widely used and efficient approach for addressing the many-electron Schrödinger equation across various domains such as physics, chemistry, and biology. However, a core challenge that persists over the long term pertains to refining the exchange-correlation (XC) approximation. This approximation significantly influences the triumphs and shortcomings observed in DFT applications. Nonetheless, a prevalent issue among XC approximations is the presence of systematic errors, stemming from deviations from the mathematical properties of the exact XC functional. For example, although both B3LYP and DM21 (DeepMind 21) exhibit improvements over previous benchmarks, there is still potential for further refinement. In this paper, we propose a strategy for enhancing XC approximations by estimating the neural uncertainty of the XC functional, named Residual XC-Uncertain Functional. Specifically, our approach involves training a neural network to predict both the mean and variance of the XC functional, treating it as a Gaussian distribution. To ensure stability in each sampling point, we construct the mean by combining traditional XC approximations with our neural predictions, mitigating the risk of divergence or vanishing values. It is crucial to highlight that our methodology excels particularly in cases where systematic errors are pronounced. Empirical outcomes from three benchmark tests substantiate the superiority of our approach over existing state-of-the-art methods. Our approach not only surpasses related techniques but also significantly outperforms both the popular B3LYP and the recent DM21 methods, achieving average RMSE improvements of 62\% and 37\%, respectively, across the three benchmarks: W4-17, G21EA, and G21IP. 
\end{abstract}

\section{Introduction}
Density Functional Theory (DFT) has gained widespread popularity as a technique for calculating electronic structures and energies, unveiling physical properties in various fields including chemistry, material science, biology, and medicine. This popularity arises from a well-accepted compromise between computational cost and accuracy, as highlighted by the consensus \cite{Burke2012dft}. While an exact functional method from electron density to energy theoretically exists and has been demonstrated \cite{Hohenberg1964dft}, the exponential time complexity in relation to the number of electrons renders practical implementation of DFT unfeasible.To address this, Kohn-Sham (K-S) formalism was introduced \cite{Kohn1965dft}, significantly reducing the computational complexity of DFT to cubic order. This achievement was realized by leveraging the concept of non-interacting electrons within a many-electron system, leading to the formulation of a set of self-consistent equations. Despite this advancement, a persistent challenge lies in the development of suitable XC approximations for the K-S equations in practical calculations \cite{Medvedev2017dft}.

In the preceding decades, three classical XC approximations have emerged. The Local Density Approximation (LDA) \cite{Kohn1965dft}, stands as the simplest and most prevalent functional for accurately describing molecular geometries. However, LDA exhibits a significant drawback by strongly overbinding molecules, leading to an overestimation of about 1 eV per bond. To address this limitation, the Generalized Gradient Approximation (GGA) \cite{Perdew1986GGA}, takes a stride by incorporating the first derivative of the density, thus improving the accuracy of chemical calculations and rectifying the overbinding issue present in LDA. Furthermore, B3LYP \cite{Becke1993hybrid} ingeniously integrates the Hartree-Fock (HF) exchange functional, amalgamating it with various local exchange and correlation functionals, such as LDA \cite{Kohn1965dft}, Becke's 1988 (B88) \cite{Becke1988B88}, Lee-Yang-Parr (LYP) \cite{Lee1988LYP}, and Vosko-Wilk-Nusair (VWN) \cite{Vosko1980VWN}. B3LYP has become a cornerstone in various domains of chemistry due to its improved accuracy. However, the drawback lies in the fact that these hybrid functional coefficients are largely manually calibrated, yielding hand-designed functionals that often fail to surpass the B3LYP \cite{Cohen2012dft}.

Recently, DeepMind 21 (DM21) \cite{Kirkpatrick2021_DM21} has introduced a novel approach that combines local LDA correlation and HF exchange functionals through a Multilayer Perceptron (MLP). This MLP predicts the local enhancement factors, essentially the combined coefficients, enabling the integration of local energies throughout space. Once trained, DM21 proves effective in Self-Consistent Field (SCF) calculations, offering precise results for intricate systems like chemical reactions. Although DM21 surpasses leading hybrid XC functionals such as B3LYP \cite{Becke1993hybrid}, M06-2X \cite{Zhao2008M062X}, and $\omega$B97X \cite{Chai2008wB97X}, all of them still have significant systematic errors compared to exact (or highly accurate) functionals, see the DM21 results in benchmark datasets like GMTKN55 \cite{Goerigk2010GMTKN55} and QM9 \cite{Kim2019QM9}. In addition, to estimate the above systematic errors, \cite{Mortensen2005Bayesian,Aldegunde2016Uncertainty} use Bayesian statistics to construct an ideal functional that combines an ensemble of existing XC functionals and a Gaussian error for uncertainty quantification.

To better capture systematic errors, we introduce a residual uncertainty framework to characterize the eXchange-Correlation (XC) functional. Drawing inspiration from uncertainty modeling in machine learning and computer vision \cite{kendall2017uncertainties} and the Bayesian error estimation of DFT \cite{Mortensen2005Bayesian}, we conceive of the XC functional as a Gaussian distribution characterized by its mean and variance. Subsequently, we develop a neural network to predict both the mean and variance, facilitating an accurate approximation of the XC functional. This neural network takes into account both local and nonlocal features of the occupied Kohn-Sham (KS) orbitals \cite{Kirkpatrick2021_DM21}. Challenges arise in achieving convergence in Self-Consistent Field (SCF) calculations for the KS equations when directly predicting the mean and variance. This sensitivity is attributed to the integration of local energies across all space. To mitigate this issue, we are inspired by the residual learning \cite{He2016resnet}, and construct the mean by amalgamating traditional XC approximations with the network predictive mean. Moreover, we regulate the predictive variance using a scaling factor to prevent excessive disruptions. Consequently, based on the aforementioned analysis, we propose a Residual Bayesian network (RBNet) designed to forecast both the mean and variance of the XC functional. It is noteworthy that our RBNet can seamlessly integrate with established traditional XC approximations (\eg, B3LYP and DM21). To confirm this, our approach exhibits substantial enhancements over both B3LYP and DM21 methodologies. Overall, our contributions can be summarized as follow:

\begin{itemize}
    \item We propose a generalized residual XC-uncertain functional for density functional theory by introducing a Residual Bayesian Network (RBNet), a novel Residual uncertainty method designed to enhance the modeling of the XC functional by forecasting both mean and variance. To the best of our knowledge, our contribution lies in pioneering the integration of neural uncertainty into the development of XC functionals.
    \item Furthermore, our approach exhibits wide applicability across various XC approximation methods. This is particularly advantageous when these methods exhibit substantial systematic errors, as our approach excels in rectifying and enhancing their performance.
    \item We comprehensively evaluate the efficacy of our methodology using four prominent public benchmarks: W4-17, G21EA, and G21IP. The experimental findings unequivocally demonstrate the remarkable superiority of our approach compared to existing exchange-correlation approximation algorithms. Crucially, our approach consistently outperforms not only the renowned B3LYP method but also the recent DM21 model, underscoring its exceptional generalization capability.
\end{itemize}

\section{Related Work}
\subsection{DFT using Deep Learning}
Density Functional Theory (DFT) has revolutionized the application of quantum mechanics to intricate challenges in chemistry, materials science, biology, and medicine. Reviews such as \cite{Cohen2012dft,Burke2012dft,Medvedev2017dft} encompass most conventional DFT methods. In this context, we focus on a comprehensive examination of eXchange-Correlation (XC) approximation techniques utilizing deep learning \cite{Pederson2022ml-dft}. The concept involves training a neural network to serve as a universal XC functional, capable of accurately reproducing both the exact XC energy and potential \cite{Schmidt20219NN-exchange-potential}. Additionally, we explore the utilization of a three-dimensional convolutional neural network designed to approach an exact XC potential \cite{Zhou202193DCNN-exchange}. Another avenue involves NeuralXC, which employs a Behler-Parrinello-style neural network architecture to generate non-local XC functionals \cite{Dick2020ml-exchange}.

The implementation of neural networks for the exchange-correlation functional holds the potential to enhance simulation precision within the framework of differentiable Kohn-Sham density functional theory \cite{Kasim2021learn-xc}. Here, we adopt machine learning algorithms to semi-automate the creation of X and XC holes via artificial neural networks \cite{Cuierrier2021xc-nn}. The act of training neural networks for the exchange-correlation functional introduces implicit regularization, significantly bolstering generalization capabilities \cite{Li2021ks-diffprogram}. Furthermore, by applying a machine learning model adhering to physical asymptotic constraints, the XC functional is constructed by separating it into exchange (X) and correlation (C) components \cite{Nagai2022ml-exchange}. To enhance the strongly constrained and appropriately normed SCAN functional, a deep neural network is ingeniously designed, leveraging precise constraint adherence and appropriate norms \cite{Pokharel2022NN-exchange}.

In recent developments, DM21 employs a multilayer perceptron (MLP) to predict local enhancement factors of local LDA correlation and HF exchange functionals, particularly for the integration of local energies throughout space \cite{Kirkpatrick2021_DM21}. A derivative of this approach, known as many-body MB-DM21 potential, finds application in simulating liquid water across varying temperatures and ambient pressures \cite{Palos2022DM21water}. Despite the considerable advancements of these machine learning methodologies compared to traditional XC methods, they still exhibit notable systematic errors in benchmarking. This paper introduces a novel strategy to mitigate these systematic errors, thereby enhancing the precision of XC functional calculations.

\section{Backgrounds}

\textbf{B3LYP} \cite{Becke1993hybrid} stands out as one of the most extensively employed exchange-correlation (XC) functionals in density functional theory (DFT) calculations. This hybrid functional amalgamates the HF exchange functional, denoted as $e^{\text{HF}}_x$, with the VWN correlation functional \cite{Vosko1980VWN}, along with the local density approximation $e^{\text{LDA}}_x$ \cite{Kohn1965dft}, the Becke 88 three-parameter exchange functional $e^{\text{GGA}}_x$ \cite{Becke1988B88}, and the LYP correlation functional $e^{\text{GGA}}_c$ \cite{Lee1988LYP}. The formula for B3LYP is defined as follows: %

\begin{equation}
\begin{aligned}
e^{\text{B3LYP}}_{xc} & = e^{\text{LDA}}_{x} + e^{\text{LDA}}_{c} + a(e^{\text{HF}}_x - e^{\text{LDA}}_x) \\
& + b(e^{\text{GGA}}_x - e^{\text{LDA}}_x) + c(e^{\text{GGA}}_c - e^{\text{LDA}}_c),
\end {aligned}
\label{eq:B3LYP}
\end{equation}
where the parameters are commonly set as $a = 0.20$, $b = 0.72$, and $c = 0.81$. Due to its affordability, versatility, and potential for enhancement, the B3LYP functional has spurred the development of composite methods like B3LYP-3c \cite{Pracht2020B3LYP-3c} and B3LYP-D3-DCP \cite{Santen2015B3LYP-D3-DCP}. Typically, B3LYP is employed alongside dispersion corrections such as DFT-D, compensating for its limited consideration of noncovalent interactions.\\
\textbf{DeepMind 21 (DM21)} \cite{Kirkpatrick2021_DM21}, is firstly developed by DeepMind in 2021, which involves training a multi-layer perceptron network to address systematic errors resulting from the violation of mathematical properties of the exact exchange-correlation (XC) functional. This approach effectively characterizes instances of artificial charge delocalization and strong correlation, outperforming conventional XC functionals like B3LYP \cite{Becke1993hybrid} and $\omega$B97X \cite{Chai2008wB97X}. The DM21 model can be described by the following formalization: %
\begin{equation}
    e^{\text{DM21}}_{xc} = a_1e^{\text{LDA}} + a_2 e^{\text{HF}} + a_3 e^{\omega\text{HF}},
    \label{eq:DM21}
\end{equation}
where the local enhancement factors $a_1$, $a_2$ and $a_3$ are obtained as the three-dimensional outputs of a multi-layer perceptron network trained on molecular data as well as on hypothetical systems with fractional charge and spin. Furthermore, an extension of DM21, known as Many-body MB-DM21 \cite{Palos2022DM21water}, applies the principles of DM21 within the many-body expansion of the energy. This extension is utilized in simulations of liquid water, considering temperature variations under ambient pressure conditions.

\begin{figure*}[bhtp]
    \centering
    \includegraphics[width=0.97\linewidth]{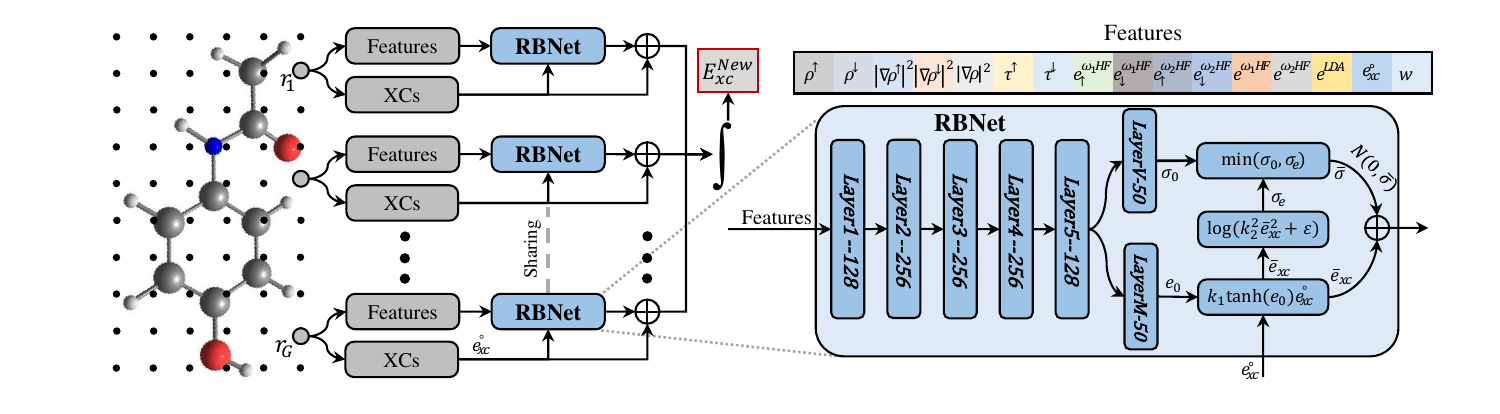}
    \caption{
    Overview of our residual XC-uncertain functional architecture. The features of the electron density are computed through sampling KS orbitals on grid points. In contrast to conventional XC functionals that produce a local XC energy density $e_{xc}^{\circ}$, our RBNet model predicts a residual energy density situated between the ideal energy density and the $e_{xc}^{\circ}$.}\label{fig:framework}
\end{figure*}

\section{Methodology}
In this section, we introduce a pioneering approach for crafting a generalized XC functional within KS-DFT calculations. Initially, we harness the concept of uncertainty to formulate a versatile structure for the XC functional while scrutinizing its inherent constraints. Subsequently, we incorporate an additional residual uncertainty to enhance the modeling of the XC functional. Finally, we elucidate the training loss associated with our proposed approach.

\subsection{Generalized XC-Uncertain Functional}
Due to the inherent difficulty in obtaining the precise XC functional, we draw inspirations from Bayesian deep learning techniques \cite{kendall2017uncertainties} and Bayesian error estimation of DFT \cite{Mortensen2005Bayesian}. In this context, we create a neural network capable of directly forecasting both the mean and variance of the XC energy density. Mathematically, we establish a generalized XC-uncertain functional represented by a Gaussian distribution characterized by mean and variance parameters: %
\begin{equation}
    e_{xc} = \widehat{e}_{xc} +\widehat{\sigma} \epsilon, \ \  [\widehat{e}_{xc},\widehat{\sigma}^2] = f(\textbf{x}(\mathbf{r}); \theta), \ \ \epsilon \sim \mathcal{N}(0,I),
    \label{eq:uncertainty}
\end{equation}
where $f(\cdot;\cdot)$ denotes a Bayesian neural network parameterized by model weights $\theta$, while $\textbf{x}(\mathbf{r})$ represents the input features at the spatial location $\mathbf{r}$. A single network can be employed to process the input $\textbf{x}(\mathbf{r})$, with its architecture designed to predict both $\widehat{e}_{xc}$ and $\widehat{\sigma}^2$ concurrently. Following the DM21 feature \cite{Kirkpatrick2021_DM21}, we adopt an 11-dimensional vector representation for the feature $\textbf{x}(\mathbf{r})$:
\begin{equation}
    \begin{aligned}
        \textbf{x}(\mathbf{r})= & [\rho_\uparrow, \rho_\downarrow, |\nabla\rho_\uparrow|^2, |\nabla\rho_\downarrow|^2, |\nabla\rho|^2,                                                                \\
                                & \tau_\uparrow, \tau_\downarrow, e_\uparrow^{\omega_1\text{HF}}, e_\downarrow^{\omega_1\text{HF}}, e_\uparrow^{\omega_2\text{HF}},e_\downarrow^{\omega_2\text{HF}}],
        \label{eq:dm21feature}
    \end{aligned}
\end{equation}
where these features are the functions of the location $\mathbf{r}$, and denoted by

\begin{itemize}
    \item The densities with spins: $\rho_\uparrow(\mathbf{r})$, $\rho_\downarrow(\mathbf{r})$
    \item The square norm of the gradient of the densities:\\
          $|\nabla\rho_\uparrow(\mathbf{r})|^2$, $|\nabla\rho_\downarrow(\mathbf{r})|^2$, $|\nabla\rho(\mathbf{r})|^2=|\nabla(\rho_\uparrow(\mathbf{r}) + \rho_\downarrow(\mathbf{r}))|^2$
    \item The kinetic energy densities with spins: $\tau_\uparrow(\mathbf{r})$, $\tau_\downarrow(\mathbf{r})$
    \item The local HF features with $\omega_1=0.4$ and $\omega_2 \xrightarrow{} \infty $:\\
          $e_\uparrow^{\omega_1\text{HF}}(\mathbf{r})$, $e_\downarrow^{\omega_1\text{HF}}(\mathbf{r})$, $ e_\uparrow^{\omega_2\text{HF}}(\mathbf{r})$ and $e_\downarrow^{\omega_2\text{HF}}(\mathbf{r})$.
\end{itemize}

Using the XC energy density $e_{xc}$ in Eq.\eqref{eq:uncertainty}, the XC energy $E_{xc}[n]$ is constructed by integrating the density $e_{xc}$:
\begin{equation}
    E_{xc}= E_{xc}^{\text{U}}+ \sigma\epsilon, \epsilon \sim \mathcal{N}(0,1),
    \label{eq:XC1}
\end{equation}
where $\rho(\mathbf{r})$ is the electron density, $E_{xc}^{\text{U}}[\rho]=\int\rho(\mathbf{r})\widehat{e}_{xc}(\mathbf{r}) d \mathbf{r}$ $=\sum_{\mathbf{r}}\rho(\mathbf{r})\overline{e}_{xc}(\mathbf{r})$ and $\sigma[\rho]=\int\rho(\mathbf{r})\widehat{\sigma}(\mathbf{r})\epsilon d \mathbf{r}=$ $\sum_{\mathbf{r}}\rho(\mathbf{r})\widehat{\sigma}(\mathbf{r})\epsilon=\sigma\epsilon$. Here, the integral represents the sum of the function on all grand points. To learn this general Uncertainty of the XC functional, we employ a minimization objective:
\begin{equation}
    \mathcal{L}_{\text{U}}=\frac{1}{N}\sum_i\frac{1}{2}\sigma_i^2\|E_{xc,i}^{\text{U}}-E_{xc,i}^{\ast}\|^2+\frac{1}{2}\log \sigma_i^2,
    \label{eq:uloss}
\end{equation}
where $N$ represents the total count of reactions, where $i$ signifies the $i$th individual reaction. $E_{xc,i}^{\text{U}}$ corresponds to the predicted total exchange-correlation energy change for the reaction (difference between product and reactant exchange-correlation energies), while $E_{xc,i}^{\ast}$ stands for a precise exchange-correlation energy for the reaction obtained from literature total energy calculations.

However, learning both $\widehat{e}_{xc}$ and $\widehat{\sigma}^2$ directly poses challenges due to integration, resulting in two distinct issues. Firstly, the sensitivity of $\widehat{e}_{xc}$ to loss convergence is heightened by the stochastically initialized network parameters. Secondly, the accumulation of variance frequently triggers loss explosion. To address these dilemmas, we introduce a concept of residual uncertainty in the subsequent subsection.

\subsection{Residual XC-Uncertain Functional}
To address the aforementioned issues, we are inspired by the residual learning \cite{He2016resnet} and introduce a concept of residual uncertainty to enhance the modeling of XC-Uncertain energy density, as depicted in Fig.\ref{fig:framework}. Our core concept involves learning a residual density situated between the theoretical energy density $\widehat{e}_{xc}$ and the conventional energy density $e_{xc}^{\circ}$. Additionally, we employ the established energy density $e_{xc}^{\circ}$ to impose constraints on the residual mean and variance, effectively preventing them from causing instability. Consequently, we define the residual XC-uncertain energy density as follows:
\begin{equation}
\begin{aligned}
    {e_{xc}=e_{xc}^{\circ}+\underbrace{\overline{e}_{xc}+\overline{\sigma}\epsilon}_{\text{RBNet}}, \epsilon \sim \mathcal{N}(0,I)},\\
    {[\overline{e}_{xc},\overline{\sigma}] = g(\textbf{y}(\mathbf{r}),e_{xc}^{\circ}; \phi), }
    \label{eq:residualuncertainty}
\end{aligned}
\end{equation}
where $e_{xc}^{\circ}$ denotes any conventional XC energy density method (in this paper, we primarily consider B3LYP ($e_{xc}^{\circ}=e_{xc}^{\text{B3LYP}}$) \cite{Becke1993hybrid} and DM21 ($e_{xc}^{\circ}=e_{xc}^{\text{DM21}}$) \cite{Kirkpatrick2021_DM21}, although numerous other methods are also applied). The function $g(\cdot;\cdot)$ represents a residual Bayesian neural network characterized by model weights $\phi$ along with two scaling factors $k_1$ and $k_2$, while $\textbf{x}(\mathbf{r})$ stands for the input features at location $\mathbf{r}$.

\textbf{Features.} Here, we extend the DM21 feature from 11-dimensional vector to 16-dimensianl vector:
\begin{equation}
    \textbf{y}(\mathbf{r})= [\textbf{x}(\mathbf{r}),e^{\omega_1\text{HF}}, e^{\omega_2 \text{HF}}, e^{\text{LDA}},e_{xc}^{\circ},w],
    \label{eq:newfeature}
\end{equation}
where $\textbf{x}(\mathbf{r})$ is defined in Eq.\eqref{eq:dm21feature}, and others are denoted by
\begin{itemize}
    \item The local HF energy with $\omega_1=0.4$ and $\omega_2 \xrightarrow{} \infty $, $e^{\omega_1 \text{HF}}=e_\uparrow^{\omega_1\text{HF}}(\mathbf{r})+e_\downarrow^{\omega_1\text{HF}}(\mathbf{r})$, $e^{\omega_2\text{HF}}=e_\uparrow^{\omega_2\text{HF}}(\mathbf{r})+e_\downarrow^{\omega_2\text{HF}}(\mathbf{r})$,
    \item The LDA energy, $e^{\text{LDA}}(\mathbf{r})$, any known energy $e_{xc}^{\circ}(\mathbf{r})$, and $w$ is the weight of the grand point.
\end{itemize}

\textbf{Residual Bayesian Network (RBNet).} To overcome the difficulty of learning the ideal $\widehat{e}_{xc}$ in the Eq.\eqref{eq:uncertainty}, we adopt the conventional XC energy density $e_{xc}^{\circ}$ as a reference point. This enables us to model a \textit{residual XC energy density} $\overline{e}_{xc}$:
\begin{equation}
    \overline{e}_{xc}=\widehat{e}_{xc}-e_{xc}^{\circ}.\label{eq:residualXC0}
\end{equation}

Subsequently, we introduce a residual Bayesian network denoted as $g(\textbf{y}(\mathbf{r}),e_{xc}^{\circ};\phi)$ to simultaneously learn both $\overline{e}_{xc}$ and $\overline{\sigma}^2$ using our novel feature $\textbf{y}(\mathbf{r})$. As illustrated in Figure\ref{fig:framework}, the RBNet comprises a unified network structure featuring two heads for the prediction of temporary values $e_0$ and $\sigma_0$. To ensure the realism of XC energy densities, we formulate $\overline{e}_{xc}$ and $\overline{\sigma}^2$ as follows:
\begin{equation}
    \begin{aligned}
        \overline{e}_{xc} & =k_1\tanh{(e_0)}e_{xc}^{\circ},     \\
        \overline{\sigma} & = \min\{\sigma_0, \sigma_{e}\}, \ \
        \sigma_{e}=\log(k_2^2\overline{e}_{xc}^2+\varepsilon),
        \label{eq:residualXC1}
    \end{aligned}
\end{equation}
where $\varepsilon$ is introduced as a small constant to prevent $\overline{e}_{xc}$ from becoming excessively small. The scaling factors $0<k_1<2$ and $0<k_2<2$ play a role in shaping the model's behavior. Specifically, we set $\varepsilon=10^{-4}$ and $k_1=k_2=1$.

\begin{table*}
    \centering
    \resizebox{0.97\linewidth}{!}{
        \begin{tabular}{c|ccc|ccc|ccc}
            \Xhline{1.3pt}
            \multirow{2}*{Methods} & \multicolumn{3}{c|}{W4-17} & \multicolumn{3}{c|}{G21EA} & \multicolumn{3}{c}{G21IP}  \\
            \cline{2-10}
                                   & RMSE $(\downarrow)$          & MAE $(\downarrow)$         & MAD $(\downarrow)$        & RMSE $(\downarrow)$ & MAE $(\downarrow)$ & MAD $(\downarrow)$ & RMSE $(\downarrow)$ & MAE $(\downarrow)$ & MAD $(\downarrow)$ \\
            \hline
            SCAN-D3(BJ)            &
            6.32                   &
            4.36                   &
            4.63                   &
            4.02                   &
            3.48                   &
            3.50                   &
            5.57                   &
            4.53                   &
            4.53  \\
            M06-2X-D3(0)           &
            5.75                   &
            3.66                   &
            4.20                   &
            2.28                   &
            1.94                   &
            1.84                   &
            3.97                   &
            2.86                   &
            2.80     \\
            $\omega$B97X-V         &
            4.97                   &
            2.77                   &
            3.65                   &
            2.31                   &
            1.92                   &
            1.85                   &
            3.83                   &
            3.16                   &
            3.21   \\
            MN15-D3(BJ)            &
            6.19                   &
            4.76                   &
            4.78                   &
            \underline{1.89}       &
            \underline{1.47}       &
            \textbf{1.21}          &
            4.12                   &
            3.16                   &
            3.15    \\
            \hline
            B3LYP-D3(BJ)           &
            6.05$\pm$2.9E-8        &
            4.68$\pm$4.4E-8        &
            2.94$\pm$4.4E-8        &
            6.31$\pm$4.0E-11       &
            5.04$\pm$9.7E-10       &
            3.45$\pm$5.4E-10       &
            5.83$\pm$1.1E-3        &
            4.09$\pm$1.2E-13       &
            3.67$\pm$2.8E-13   \\

            B3LYP+RBNet            &
            \makecell[c]{\textbf{1.95$\pm$4.9E-8}   \checkmark \\
            +67.77\%}              &
            \makecell[c]{\underline{1.29$\pm$1.2E-7} \checkmark\\
            +72.44\%}              &
            \makecell[c]{\textbf{1.30$\pm$1.4E-7}   \checkmark \\
            +55.78\%}              &
            \makecell[c]{\textbf{1.67$\pm$9.5E-4}   \checkmark\\
            +73.53\%}              &
            \makecell[c]{\textbf{1.38$\pm$3.1E-4}   \checkmark\\
            +72.62\%}              &
            \makecell[c]{1.35$\pm$2.4E-4            \checkmark\\
            +60.87\%}              &
            \makecell[c]{\textbf{2.60$\pm$3.4E-3}   \checkmark\\
            +55.40\%}              &
            \makecell[c]{\textbf{1.97$\pm$1.2E-3}   \checkmark\\
            +51.83\%}              &
            \makecell[c]{\textbf{1.79$\pm$9.6E-4}   \checkmark\\
            +51.23\%}\\
            \hline
            \text{DM21}            &
            3.41$\pm$1.3E-2        &
            2.36$\pm$1.3E-2        &
            2.13$\pm$4.9E-3        &
            2.66$\pm$7.1E-7        &
            2.09$\pm$1.9E-6        &
            1.37$\pm$8.4E-3        &
            9.97$\pm$1.9E-2        &
            5.05$\pm$2.9E-2        &
            6.54$\pm$7.4E-2\\
            \text{DM21+RBNet}      & \makecell[c]{\underline{2.45$\pm$1.2E-5}    \checkmark\\ +28.15\%} & \makecell[c]{1.74$\pm$5.5E-4                \checkmark\\ +16.75\%} & \makecell[c]{\underline{1.32$\pm$1.3E-5}    \checkmark\\ +38.03\%} & \makecell[c]{2.34$\pm$2.6E-3                \checkmark\\ +12.03\%} & \makecell[c]{1.74$\pm$5.5E-4                \checkmark\\ +16.75\%} & \makecell[c]{\underline{1.33$\pm$9.3E-4}   $\star$\\ +2.92\%}     & \makecell[c]{\underline{2.86$\pm$3.0E-3}    \checkmark\\ +71.31\%} & \makecell[c]{\underline{2.01$\pm$7.6E-4}    \checkmark\\ +60.20\%} & \makecell[c]{\underline{2.10$\pm$1.3E-3}    \checkmark\\ +67.89\%} \\
            \Xhline{1.3pt}
        \end{tabular}
        }
    \caption{    
    State-of-the-art performance (kcal/mol) by our RBNet (residual XC-uncertain functional) on W4-17, G21EA and G21IP datasets. We can see that our RBNet outperforms both B3LYP and DM21 significantly, validating its robustness and generalizability. The \textbf{bold} and {\underline{underline}} font indicate the best and second-best outcomes. The standard deviation is presented in scientific notation. The symbol $\checkmark$ denotes successful passage of Welch’s $t$-test at a significance level of 0.05, whereas $\star$ indicates successful passage at a significance level of 0.10.}\label{tab:results}
\end{table*}

\subsection{Training Loss}
Before presenting the training loss, we first show the energy through the XC energy density in Eq.\eqref{eq:residualuncertainty}, referred to as residual uncertainty, which is subsequently determined via the following integral:
\begin{equation}
    E_{xc}= E_{xc}^{\circ}+E_{xc}^{\text{RU}}+ \sigma\epsilon, \epsilon \sim \mathcal{N}(0,I),\label{eq:residualXC2}
\end{equation}
where the terms are defined as follows: $E_{xc}^{\circ}[\rho]=$ $\int \rho(\mathbf{r})e_{xc}^{\circ}(\mathbf{r}) d \mathbf{r}=\sum_{\mathbf{r}}\rho(\mathbf{r})e_{xc}^{\circ}(\mathbf{r})$, $E_{xc}^{\text{RU}}[\rho]=\int\rho(\mathbf{r})\overline{e}_{xc}(\mathbf{r}) d \mathbf{r}=\sum_{\mathbf{r}}\rho(\mathbf{r})\overline{e}_{xc}(\mathbf{r})$ and $\sigma[\rho]=\int\rho(\mathbf{r})\overline{\sigma}(\mathbf{r})\epsilon d \mathbf{r}=\sum_{\mathbf{r}}\rho(\mathbf{r})\overline{\sigma}(\mathbf{r})\epsilon=\sigma\epsilon$.

We adopt a Gaussian likelihood to model the residual uncertainty, which leads to a minimization objective, $\mathcal{L}_{\text{RBNet}} = \frac{1}{N}\sum_i\frac{1}{2}\sigma_i^2|E_{xc,i}^{\circ} - E_{xc,i}^{\ast} + E_{xc,i}^{\text{RU}}|^2 + \frac{1}{2}\log \sigma_i^2$, where $E_{xc,i}^{\circ}$ signifies the exchange-correlation energy output from the traditional method for the total reaction (product minus reactant exchange-correlation energy), $E_{xc,i}^{\text{RU}}$ represents our model’s prediction, and $E_{xc,i}^{\ast}$ is defined in Eq.\eqref{eq:uloss}. Following the approach \cite{kendall2017uncertainties}, we steer the network to predict the logarithm of the variance, $s_i := \log \sigma_i^2$:
\begin{equation}
    \mathcal{L}_{\text{RBNet}}= \frac{1}{N}\sum_i\frac{1}{2}\text{exp}(-s_i)\|E_{xc,i}^{\circ}-E_{xc,i}^{\ast} +E_{xc,i}^{\text{RU}}\|^2+\frac{1}{2}s_i.
    \label{eq:loglinkelihood}
\end{equation}
\begin{figure*}[ht]
    \centering
    \includegraphics[width=1.87\columnwidth]{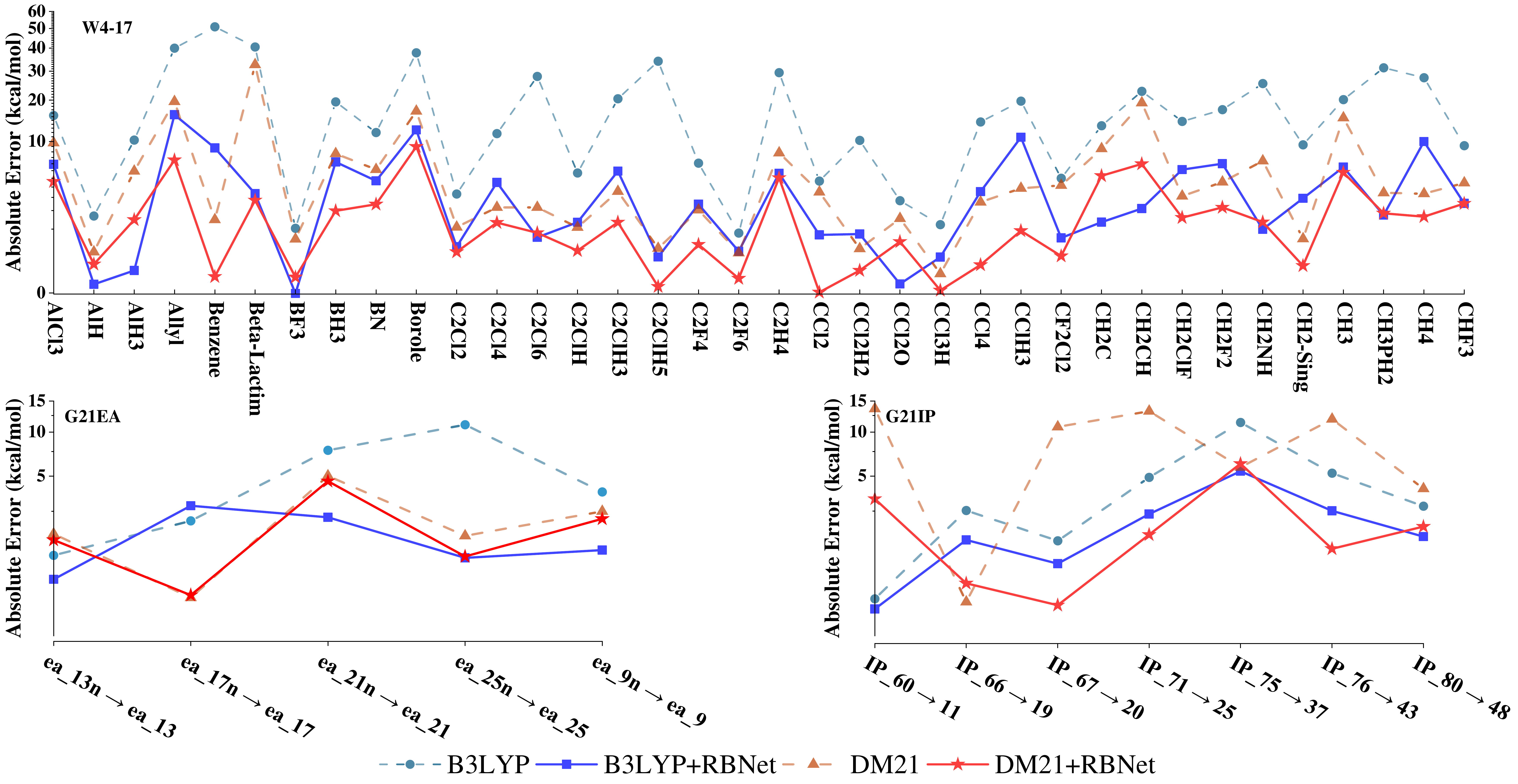}
    \caption{    
    Absolute errors (in kcal/mol) relative to the accurate energy computed from literature.}
    \label{fig:absoluteerror}
\end{figure*}

\section{Experiments}
In this section, we assess the efficacy of our methods using chemical reactions across three well-known datasets: W4-17, G21EA, and G21IP. We present the details of the experimental setup, highlight the key findings, and conduct a comprehensive parameter analysis, all of which collectively underscore the effectiveness and superiority of our RBNet.

\subsection{Experimental Setup}
\textbf{Datasets.} We consider three subsets, W4-17, G21EA and G21IP within the GMTKN55 dataset \cite{Goerigk2010GMTKN55}. The remaining portion of the analysis was performed using our proprietary single-atom energy dataset known as Atom.\textbf{W4-17} \cite{Karton2017W417} is a comprehensive collection of total atomisation energies for 200 species. These energies describe the transformational energy that occurs when one mole of a species is completely dissociated into its individual constituent atoms in the gas phase. Put simply, the total atomisation energy signifies the energy change associated with breaking all bonds within a species and converting all its atoms into isolated entities. Furthermore, \textbf{G21EA} \cite{Curtiss1991G21EA} encompasses 25 reactions involving adiabatic electron affinities, while \textbf{G21IP} \cite{Curtiss1991G21EA} incorporates 36 reactions associated with adiabatic ionisation potentials. These two datasets capture the energy changes that occur as species either gain or lose a single electron. Additionally, we computed data for isolated atoms that are implicated within these datasets, contributing to the control of training outcomes.

\textbf{Partition Settings.} We partition the datasets into training, validation, and test sets using a 6:2:2 ratio. All related atoms are included in the training set, encompassing both individual atoms and molecules consisting solely of single elements found within the dataset. Subsequently, we address each atom type based on descending atomic numbers. Should a given atom not yet be represented in the training set, a molecule containing that atom is randomly selected and added to the training set. This process ensures the inclusion of molecules featuring atom counts of 2, 3, 4, 5, 6, and those exceeding 6. Lastly, the remaining molecules are distributed randomly among the training, validation, and test sets until the desired dataset sizes are achieved.

\textbf{Implemented details.}
Our RBNet is a single network with two heads and its structure is $16-128-256-256-256-128-\left\{\begin{array}{ll}
        50-1 \\
        50-1 \\
    \end{array}
    \right.$. We train the RBNet using the loss in the Eq.\eqref{eq:loglinkelihood}. The parameters are set to $k_1=1$, $k_2=1$. For model optimization, we use the SGD optimizer and set an initial learning rate of $1e^{-3}$ with a cosine annealed schedule. We train all methods --- epochs. We implement our model using Python 3.7.16 with Pytorch 1.12.1 on a server equipped with Intel(R) Xeon(R) Gold 6230R CPU and GeForce RTX 3090. 

\textbf{Evaluation metrics.}
For all datasets, we utilize the Root Mean Squared Error (RMSE), Mean Absolute Error (MAE), and Mean Absolute Deviation (MAD) as evaluation criteria to assess the performance of all methods. RMSE quantifies the square root of the average squared disparity between predicted and actual values within a dataset. MAE gauges the mean absolute disparity between predicted and actual values in a dataset. A lower MAE indicates a better fitting model for the dataset. MAD holds particular value as a metric due to its resilience to outliers compared to alternative measures of dispersion like standard deviation and variance.

\textbf{Compared methods.}
To validate the viability of our approach, we conduct a comparative analysis against six state-of-the-art techniques as outlined below:
\begin{itemize}
    \item SCAN-D3(BJ) \cite{Sun2015SCAN} stands as a meta-generalized-gradient approximation (meta-GGA), compliant with all 17 exact constraints.
    \item M06-2X-D3(0) \cite{Zhao2008M062X} introduces a highly nonlocal functional featuring double the nonlocal exchange content (2X).
    \item MN15-D3(BJ) \cite{Yu2016MN15} presents a hybrid XC functional that amalgamates nonlocal Hartree–Fock exchange energy, nonseparable local exchange–correlation energy, and additional correlation energy.
    \item $\omega$B97X-V \cite{Mardirossian2014wB97X-v} boasts a configuration with 7 linear parameters (2 for local exchange, 4 for local correlation, and 1 for short-range exact exchange) and 3 nonlinear parameters (1 for range-separation and 2 for non-local correlation), totaling 10 optimized parameters.
    \item B3LYP-D3(BJ) \cite{Becke1993hybrid}, one of the most prevalent hybrid XC functionals employed in DFT calculations, is defined by Eq.\eqref{eq:B3LYP}.
    \item DM21 \cite{Kirkpatrick2021_DM21} predicts three local enhancement factors, blending the local density approximation $e_{xc}^{\text{LDA}}$, local Hartree Fock exchange energy densities $e^{\omega_1\text{HF}}$, and $e^{\omega_2\text{HF}}$ as expressed in Eq.\eqref{eq:DM21}.
\end{itemize}
In addition, our method interfaces with two established techniques. B3LYP+RBNet combines our residual uncertain XC functional with B3LYP by adopting it as the conventional XC functional. DM21+RBNet further merges our residual uncertain XC functional with DM21 as the conventional XC functional. Note that D3(BJ) and D3(0) are the basis of the wave function in the DFT calculation.

\begin{figure*}[ht]
    \centering
    \includegraphics[width=1.92\columnwidth]{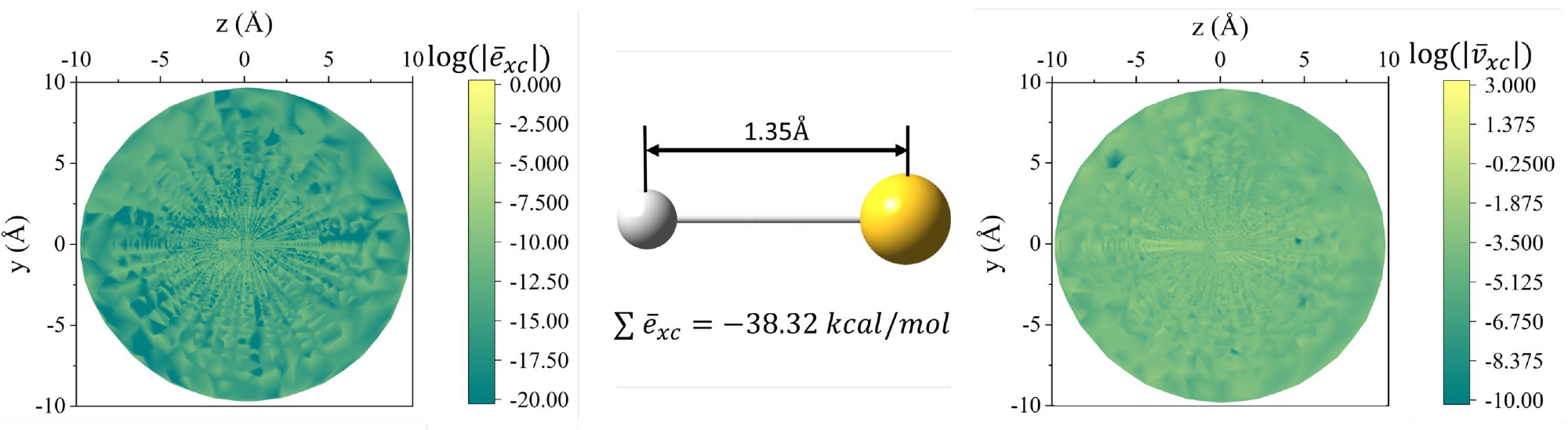}
    \caption{
    Spatial distribution of the residuals learned by our RBNet using B3LYP for sulfanyl ($HS$) in the G21EA dataset. The left and right panels show the spatial distribution of the residuals for $\log(\overline{e}_{xc})$ and for $\log(\overline{v}_{xc})$, while the middle panel displays the molecular geometry of $HS$. The highest local residuals are in the region around the nuclei and at the boundaries.}\label{fig:spacialdistributionB3LYP}
\end{figure*}
\begin{figure*}[ht]
    \centering
    \includegraphics[width=1.92\columnwidth]{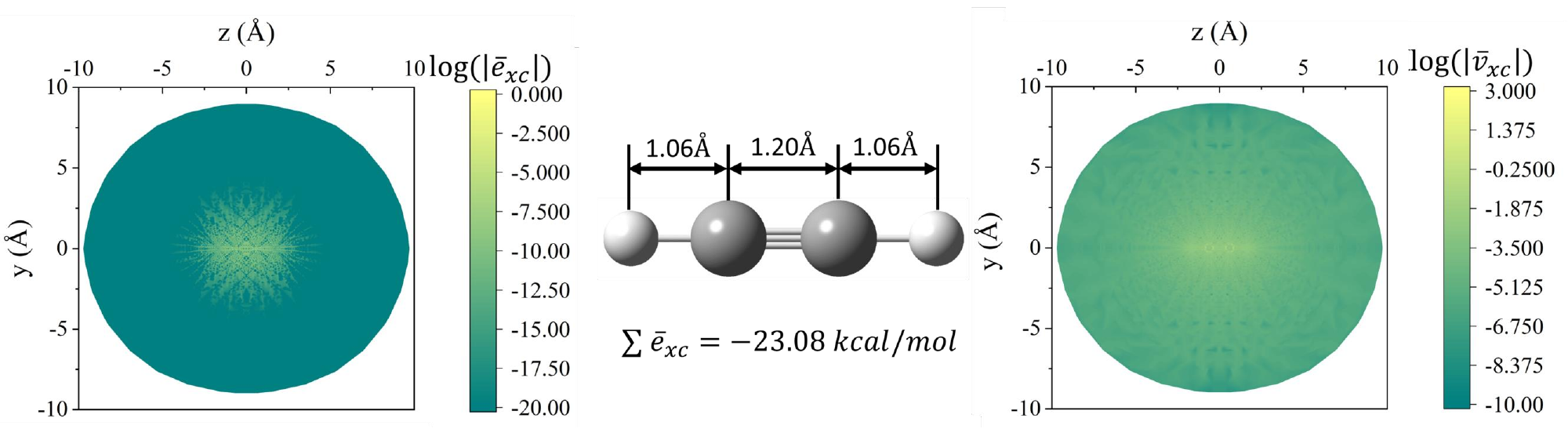}
    \caption{
    Spatial distribution of the residuals learned by our RBNet using DM21 for ethyne ($C_2H_2$) in the W4-17 dataset. The left and right panels show the spatial distribution of the residuals for $\log(\overline{e}_{xc})$ and for $\log(\overline{v}_{xc})$, while the middle panel displays the molecular geometry of $C_2H_2$. The highest local residuals are in the region around the nuclei and at the boundaries.}\label{fig:spacialdistributionDM21}
\end{figure*}

\subsection{Main Results}
\textbf{Error Results.} Table \ref{tab:results} presents the error results. Our RBNet, a residual uncertain XC functional, exhibits remarkable improvements across all datasets compared to the six XC functionals. Further insights are as follows.

Firstly, it is evident that both our B3LYP+RBNet and DM21+RBNet functionals outperform all state-of-the-art XC functionals in terms of RMSE, MAE, and MAD on the W4-17 and G21IP datasets. For instance, B3LYP+RBNet surpasses DM21 by over 1.27 kcal/mol on W4-17 and $\omega$B97X-V by more than 1.29 kcal/mol on G21IP. Moreover, our approach demonstrates comparability to MN15-D3(BJ) on the G21EA dataset. Secondly, B3LYP+RBNet exhibits superior capabilities in comparison to our baseline B3LYP, showcasing substantial RMSE improvements of 4.10, 3.50, and 2.77 kcal/mol on the three datasets, respectively. Similar observations apply to DM21+RBNet. This emphasizes RBNet's potential for generalization to other related methods. Fig.\ref{fig:absoluteerror} visually depicts the errors (in kcal/mol) relative to the accurate energies computed from literature for each test reaction.

\textbf{Visualization.} We utilize visualization techniques to analyze the spatial distribution of residuals acquired through our RBNet on both the XC energy and the XC potential. To gain a comprehensive understanding of our RBNet's performance, we focus on the XC energy and potential residuals. The results are illustrated in Figure\ref{fig:spacialdistributionB3LYP}, which shows the local residual distribution between B3LYP and B3LYP+RBNet in the sulfanyl ($HS$) molecule for $\log(\overline{e}_{xc})$ and $\log(\overline{v}_{xc})$. Likewise, Figure\ref{fig:spacialdistributionDM21} shows the distribution of local residuals between DM21 and DM21+RBNet in the ethyne ($C_2H_2$) molecule for $\log(\overline{e}_{xc})$ and $\log(\overline{v}_{xc})$. The spatial distribution unmistakably reveals elevated residual values around the nuclei and along the molecular boundaries. Additionally, a discernible spatial symmetry is observed in the distribution, centered around the molecular structure.
\begin{table}[t]
    \centering
    \resizebox{0.97\linewidth}{!}{
        \begin{tabular}{c|ccc|ccc|ccc}
            \Xhline{1.3pt}
            \multirow{2}*{Features}  & \multicolumn{3}{c|}{W4-17} & \multicolumn{3}{c|}{G21EA} & \multicolumn{3}{c}{G21IP}                                                                                                 \\
            \cline{2-10}  & RMSE & MAE & MAD & RMSE & MAE & MAD & RMSE & MAE & MAD \\
            \hline
            $\textbf{x}(\mathbf{r})$ & 2.35                         & 3.27                       & 2.16                      & 3.51          & 3.23          & 2.75          & 3.47          & 2.82          & 2.92          \\
            $\textbf{y}(\mathbf{r})$ & \textbf{1.95}                & \textbf{1.29}              & \textbf{1.30}             & \textbf{2.82} & \textbf{1.38} & \textbf{1.35} & \textbf{2.54} & \textbf{1.92} & \textbf{1.75} \\
            \Xhline{1.3pt}
        \end{tabular}}
    \caption{    
    Ablation study of the features $\textbf{y}(\mathbf{r})$ and $\textbf{x}(\mathbf{r})$.}\label{tab:ablationxy}
\end{table}
\begin{table}[t]
    \centering
    \resizebox{0.97\linewidth}{!}{
        \begin{tabular}{c|ccc|ccc|ccc}
            \Xhline{1.3pt}
            \multirow{2}*{Methods} & \multicolumn{3}{c|}{W4-17} & \multicolumn{3}{c|}{G21EA} & \multicolumn{3}{c}{G21IP}                                                                                                 \\
            \cline{2-10}
                                   & RMSE                         & MAE                        & MAD                       & RMSE          & MAE           & MAD           & RMSE          & MAE           & MAD           \\
            \hline
            B3LYP (Baseline)       & 6.05                         & 4.68                       & 2.94                      & 6.32          & 5.04          & 3.45          & 5.31          & 4.02          & 3.66          \\
            +ResNet                & 3.94                         & 2.71                       & 2.27                      & 3.27          & 2.62          & 1.71          & 3.74          & 2.97          & 2.74          \\
            +RBNet                 & \textbf{1.95}                & \textbf{1.29}              & \textbf{1.30}             & \textbf{2.82} & \textbf{1.38} & \textbf{1.35} & \textbf{2.54} & \textbf{1.92} & \textbf{1.75} \\
            \Xhline{1.3pt}
        \end{tabular}}
    \caption{
    Ablation study of the ResNet and our RBNet.}\label{tab:ablationRBNet}
\end{table} 

\textbf{Ablation Study.} Since we design a new feature $\textbf{y}(\mathbf{r})$ and propose the RBNet, we conduct two ablation studies to validate their effectiveness, utilizing B3LYP as the baseline. First, comparing our feature $\textbf{y}(\mathbf{r})$ to the feature $\textbf{x}(\mathbf{r})$ from DM21, Table \ref{tab:ablationxy} illustrates that, with B3LYP+RBNet, our $\textbf{y}(\mathbf{r})$ outperforms $\textbf{x}(\mathbf{r})$ across all three datasets. Second, by incorporating a ResNet without variance and our proposed RBNet to learn XC functionals, denoted as B3LYP+ResNet and B3LYP+RBNet, respectively, Table \ref{tab:ablationRBNet} reports the results to show that our RBNet exhibits the capability to capture superior XC functionals, thus enhancing the accuracy of DFT.

\begin{figure}[t]
    \centering
    \includegraphics[width=0.97\columnwidth]{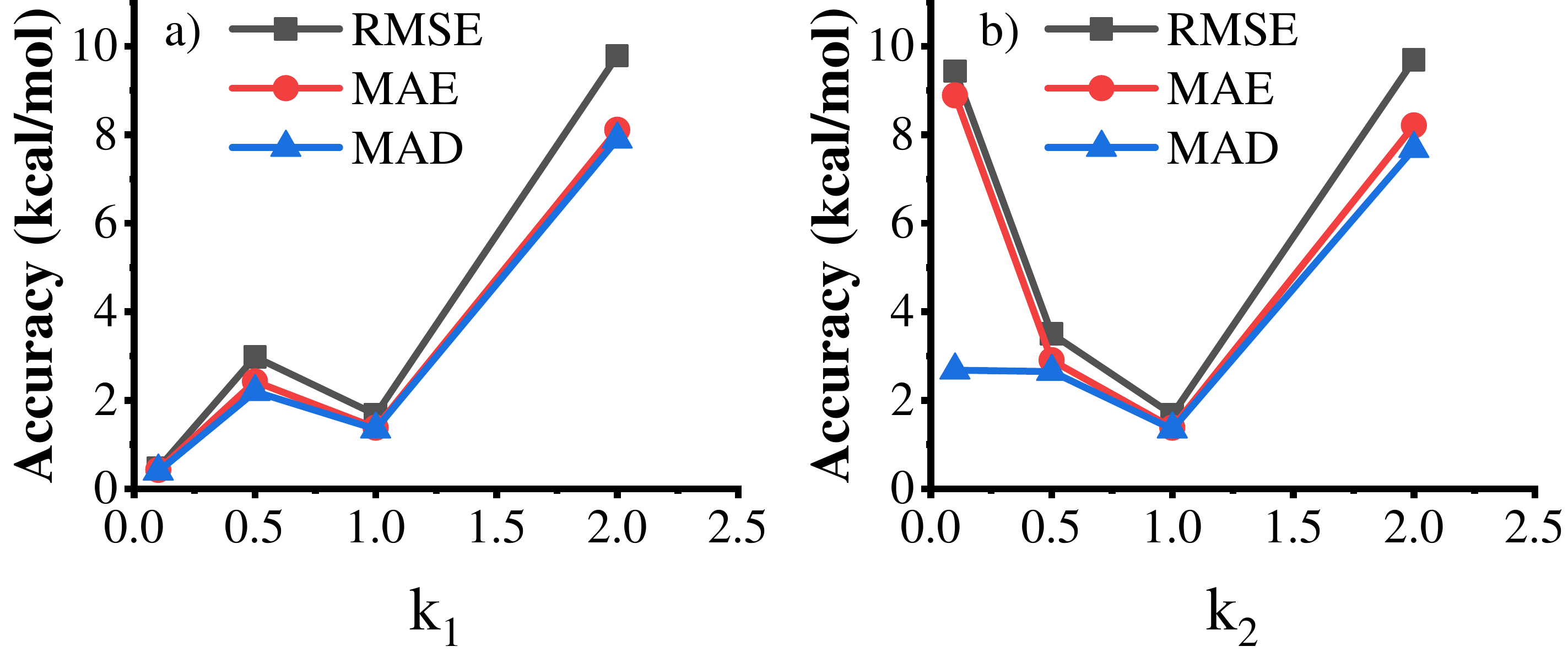}
    \caption{
    Parameter analysis of two parameters $k_1$ and $k_2$. Left shows the results of $k_1$ varied from $0.1$ to $2.0$ when $k_2=1$, while right showcase the results of $k_2$ varied from $0.1$ to $2.0$ when $k_1=1$.}\label{fig:parameteranalysis}
\end{figure}

\textbf{Parameter Analysis.}
Our network incorporates two essential hyperparameters: $k_1$ and $k_2$. The role of $k_1$ is to define the upper limit for the mean of the correction, while $k_2$ serves to set the upper limit for the variance of the correction. When they increase or decrease, the mean and variance will get bigger or smaller, resulting in an explosion or no-change of the sum energy. To identify optimal values, we employ a control variable approach. As depicted in Fig.\ref{fig:parameteranalysis}, during practical implementation, we ascertain that $k_1=1$ and $k_2=1$.

\section{Conclusion}
In this paper, we introduced a novel uncertain format for the XC Functional, referred to as the General XC-Uncertain Functional, which involves training a single neural network to determine both the mean and variance. Nevertheless, directly learning these parameters is challenging due to issues such as non-convergence in integration results or the potential for training loss explosion. To overcome this, we proposed a solution involving a residual XC-uncertain functional. It was achieved by introducing a residual Bayesian network, designed to learn the residual approximation between the ideal (ground-truth) functional and the conventional XC functional. Our experimental findings demonstrated the effectiveness of our approach, showcasing its superiority over existing state-of-the-art XC functionals, like B3LYP and DM21.

Moreover, there are two directions for investigation: 1) \textit{Enhancing Accuracy:} Despite significant strides in minimizing the disparity between our predictions and the ground-truth values, additional efforts are imperative to refine the precision of our predictions further. 2) \textit{Generalization:} Our method is inherently adaptable for extension to accommodate other XC functionals. This underscores the versatility and potential broader applicability of our approach.

\section{Acknowledgements} Jun Li was partially supported by the National Science Fund of China under Grant Nos. 62072242.
\bibliography{aaai25}
\end{document}